\documentclass{article}
\usepackage{epsf}
\usepackage[dvips]{graphics}

\begin{document}

\leftline{{\bf Paper number:} E090}

\vskip 0.5in
\leftline{\bf Title:}
\par
Conformal lattice of magnetic bubble domains in garnet film

\vskip 0.5in
\leftline{\bf Authors:}
\par
M.~Gutowski, K.~Piotrowski, M.~Gutowska and A.~Szewczyk

\bigskip
\leftline{{\bf Address} (valid for all authors):}
\par
Institute of Physics, Polish Academy of Sciences
\par
Al. Lotnik\'ow 32/46, PL~02--668 Warsaw, Poland

\vskip 1in
\leftline{\bf Abstract:}

%\begin{abstract}
We report experimental observations of magnetic bubble domain arrays
with no apparent translational symmetry. Additionally the results of
comparative numerical studies are discussed.  Our goal is to present
experimental evidence for {\em natural\/} occurrence of conformal
structures.

%\end{abstract}

\bigskip
\leftline{\bf Keywords:}
\par
conformal matter, domain pattern, garnets -- thin films, magnetic
bubbles, triangular lattice

\vskip 1in
\leftline{\bf Corresponding author:}
\par
Marek W. Gutowski
\par
Institute of Physics, Polish Academy of Sciences
\par
PL--02--668 Warsaw (Warszawa)
\par
POLAND
\par
e-mail: gutow@ifpan.edu.pl
\par
tel.:  +48-22-8437001 ext. 3122
\par
fax:   +48-22-8430926

\vfil
\newpage

\baselineskip=1.5\baselineskip

%\section{Introduction}

Our work was motivated by few earlier articles \cite{firstPieran,
smallPieran, Rothen, Wojciech, KOWBAN} concerning the equilibrium
configurations of identical interacting objects subjected additionally
to strong external fields.  While majority of papers concentrate on
small-- to medium--sized clusters and their thermodynamic
\cite{mesoclu, radial} and elastic \cite{Rothen, Wojciech} properties,
the aim of this paper is to demonstrate the very existence of conformal
structures in the Nature, beyond the traditional examples of living
things, like sunflower and alike.

The trivial examples of conformal transformations include: Euclidean
"rigid movements", i.e. translations and rotations, point reflections,
as well as uniform scaling, simultaneously in all dimensions. 
Generally, the conformal image of any domain contains parts which are
stretched, compressed and rotated.  Here, we are interested in such
nontrivial conformal mappings, which do not preserve the distances
between images of originally equidistant points.

%\section{Experimental setup}

The object of our study was the thin layer of magnetic material
(garnet, with composition given by its chemical formula
(TmBi)$_{3}$(FeGa)$_{5}$O$_{12}$) deposited epitaxially on non-magnetic
substrate (GGG, $\langle111\rangle$ oriented). The sample was held at
room temperature in the uniform bias field, produced by a pair of
Helmholtz coils, directed perpendicularly to the sample plane. The
magnitude of bias field was chosen in such a~way to fall within range,
in which the so called bubble domains exist and are stable
\cite{bubble}.  Due to their size, with diameter around $10$~$\mu$m,
those domains can be easily viewed using ordinary optical microscope
equipped with two polarizers.  The Faraday effect, i.e. the difference
of refraction indices for circularly polarized light propagating in
direction parallel or anti-parallel to the magnetization vector, is
responsible for the contrast in the observed images. We have used the
white light in our investigations. Many authors \cite{bubble-arrays}
reported observations of regular triangular or hexagonal arrays of
bubble domains created in such conditions. In addition to the coils
producing the bias field, we have used another small coil, with
diameter of few $mm$, driven with AC current ($\sim\!15$~kHz,
rectangular waveform), resting on the sample, to generate and
equilibrate new domain structures.  In several cases we have also used
a~DC current flowing through the additional small coil, thus mimicking
the confinement of domains by axially symmetric parabolic potential, as
in \cite{balia}.

\begin{figure}[f]
 \epsfysize=5.614cm %5.625cm %6.123
 \epsfxsize=7.5cm
 \centerline{\epsfbox{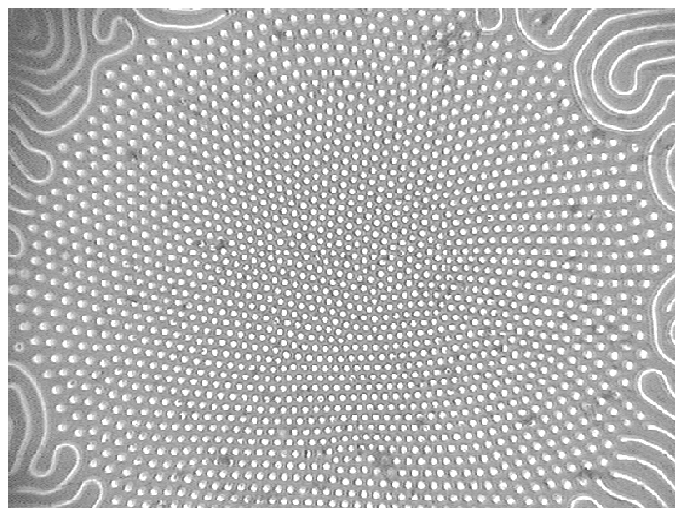}}
\caption{Microscopic image of the bubble array structure.}
\label{bubbles}
\end{figure}

The resulting picture, after sudden switching off the current in an AC
coil, was subsequently captured to the computer memory by means of
a~CCD camera, with MultiScan software by CSS Ltd.\  The obtained bitmap
file was subsequently converted to a~list of positions of the centers
of all visible bubbles and stored in text format.

%\section{Experimental data processing and results}

A single example of many generated structures is shown in
Fig.~\ref{bubbles}. The interacting bubble domains organize themselves
in a structure, which is neither periodic nor chaotic. One can easily
see that almost all bubble domains have six nearest neighbors forming
a~slightly distorted hexagon. The hexagons themselves differ in their
sizes and orientations, so there is no long range ordering in the usual
sense, i.e. the observed pattern has no translational symmetry.  On the
other hand, we may distinguish fairly regular, crystal-like regions,
which may be regarded as a~set of crossing points of three families of
arcs corresponding to the lattice lines.

The angles, at which those arcs cross, are of main interest in this
work.  We are going to show that they all are approximately equal to
$\pi/3$ ($60^{\circ}$), like in ordinary triangular (hexagonal)
lattice. This result remains true even for extremely deformed
structures, in which the ratio of distances between nearest neighbors
located in various regions of the sample reaches the value of $\sim\!3$.

The first step in experimental data processing is the triangulation,
which is performed using our own Fortran procedure, which perhaps
should be replaced by, tailored to our needs, standard Delaunay
triangulation. As a result of this step we obtain the list of nearest
neighbors for each bubble domain. This way we are able to identify the
short segments approximating the lattice lines, each segment consisting
of two straight pieces sharing common endpoint.

Two segments, with common endpoint, located at any given bubble domain,
are considered as approximating the same lattice line when the angle
between them does not exceed some predefined threshold value,
$\Theta_{t}$. The results are not very sensitive to the exact value of
$\Theta_{t}$, unless it is unreasonably high or small.  Taking
$\Theta_{t}$ too close to zero results in ignoring many connections
between lattice nodes, which are obvious for human eye.  Contrary, if
$\Theta_{t}$ is too high, then we face many misinterpretations,
especially in the vicinity of sample defects.  We have checked, by
careful inspection of the results obtained with $\Theta_{t}$ in range
$\left[ 0^{\circ}, 35^{\circ}\right]$ with step of\ $0.5^{\circ}$, that
almost identical results are produced with $\Theta_{t} \in
\left[7^{\circ}, 20^{\circ}\right]$, so we adopted in subsequent
calculations $\Theta_{t} =\pi/12$,\ i.e. $15^{\circ}$ as a safe value.

In regular cases, i.e. away from the boundaries of investigated data
set, we usually find three sections of "lattice lines" crossing at any
given node. The task is to measure the angles between the true lattice
lines, which are different from those between straight segments.  To
achieve this, we make use of the notion of curvature and curvature
radius. A circle passing through each triade of consecutive points
(lattice nodes connected with segments) is constructed, together with
vector parallel to curvature radius pointing to the middle node of the
triade.  The scalar product of two such vectors, after their
normalization, gives us the cosine of the searched angle $\Theta$.  The
rest is obvious: the procedure is repeated for every node and the
results are stored in computer's memory.

\begin{figure}[f]
 \epsfxsize=7.5cm
 \centerline{\epsfbox{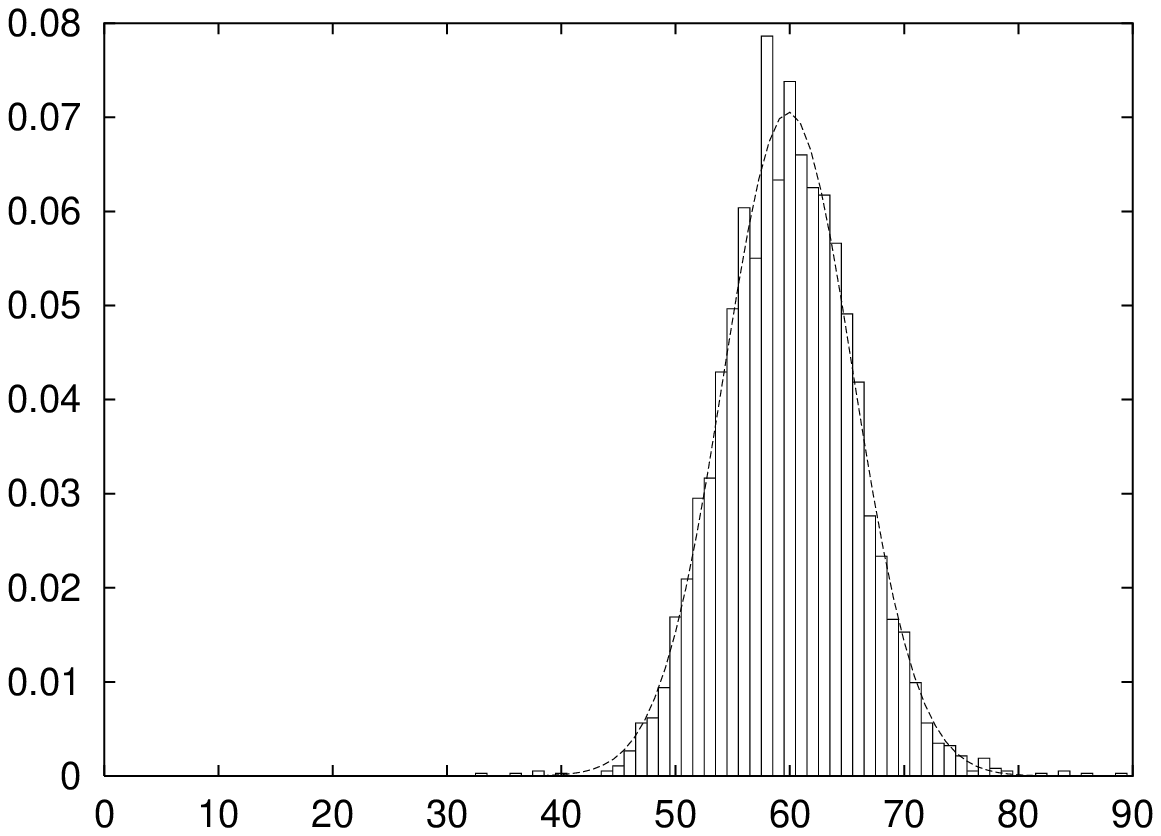}}
\caption{Distribution of crossing angles for the structure shown in
Fig.~1.  The center of the peak is located at
\protect{$59.889^{\circ}$} and its dispersion is
\protect{$5.655^{\circ}$}.  For comparison the Gaussian distribution
with the same parameters is superimposed on the original plot. The
values given here are representative for the broad range of
experimental conditions.}
\label{hist}
\end{figure}

One of many distributions of crossing angles, obtained this way, is
presented in Fig.~\ref{hist}. By broad range of experimental conditions
we mean various values of the bias field as well as the additional
field produced by small coil.

%\section{Numerical simulations}

We have also performed some numerical Monte Carlo simulations combined
with simulated annealing. In the course of simulations we have adopted
the following assumptions: {\em i\/})~the domains are identical,
non-penetrable hard disks with fixed diameter and carrying identical
magnetic moments, {\em ii\/})~each single domain is represented by
a~magnetic dipole with only two translational degrees of freedom, and
{\em iii\/})~the external field is directed perpedicularly to the
sample plane and its strength changes quadratically with position
relative to the center of available circular region, bounded with hard
wall potential.

Starting with random initial configuration the Brownian jump in
randomly chosen direction was attempted for each domain. The average
amplitude of jumps was dynamically adjusted in such a way, that
$\sim\!50$\% of them were accepted at current temperature, according to
the well known Metropolis algorithm. The temperature was slowly, but
systematically, reduced until the amplitude of jumps decreased to one
thousands of disk diameter.  The one-time decrease of temperature never
exceeded one half of the average single domain energy change
(fluctuation) in last simulation cycle. It was possible to trace the
evolution of the entire system, not only its final configuration.  In
particular, we were able to observe the collective movements of
domains, in form of outer shell rotations, previously described in
\cite{mesoclu}.  The radial diffusion \cite{radial}, although certainly
present, was not so evident.

%\section{Discussion}

We conjecture that another example of conformal matter could be the
vortex lattice in type II superconductors, which is, unfortunately,
only rarely directly observed due to the small size of each single
vortex.  We have tested our method on the scanned image of vortex
lattice taken from \cite{rose-innes} and have found results very
similar to those presented in Fig.~\ref{hist}. This, however, may be
accidental, since the lattice presented there contains only a small
number of defects, and in such case our method should produce the
histogram consisting of precisely one non-empty bin, located exactly at
$\pi/3$ ($60^{\circ}$).  Additionally, the above lattice shows almost
neglible variance of distances between nearest neighbors.

We think, that small clusters (and, perhaps, quasicrystals and
nucleation centers) might be the formidable objects for further studies
concerning conformal structures.  The conformal structures should be
created also when the interactions are purely repulsive, but the set of
objects is confined in space either by strong external field or by
hard-wall potential (we have investigated both such cases in our
computer simulations). In the latter case even the external field is no
longer necessary to create conformal structure, which may be hard to
detect, but on the other hand -- this very structure may be responsible
for increased stability of small clusters compared to their bulk
counterparts, as was very recently demonstrated for tin clusters
\cite{smclust} containing 15---30 atoms.  The cited observation may be
one of the first clear indications, that conformal matter, if this is
the case, is the qualitatively different form of condensed matter.

Our computer simulations show, that conformal matter may exist in a
nice form only as an assembly of relatively small number of individual
constituents, typically $80$--$200$ objects.  Larger sets of objects
tend to form the structures resembling those of polycrystals, no matter
how slowly the temperature is decreased during simulated
crystallization.  The role of {\em magic numbers\/}
\cite{mesoclu,balia} of the objects is not clear in those cases.  On
the other hand, it is obvious that the conformal matter cannot exist in
arbitrarily large chunks -- they would have to be unstable.  Our
calculations of structure factors for the conformal crystals,
consisting of few hundreds of objects, show that the resulting X-ray
image reflects the symmetry elements of the sample as a whole rather
than the local (six-fold) symmetry, if any.  In this respect the
conformal matter appears as an amorphous material. In conclusion, it
would be rather difficult to detects non-trivial conformal forms by
traditional X-ray (or neutron) analysis.

%\section{Conclusions}

Summarizing, we think that conformal (poly)crystals should be regarded
as another form of the condensed matter, different from well known
periodic, quasiperiodic \cite{qcryst}, modulated or incommensurate
crystals.

This work was supported in part within European Community program
ICA1-CT-2000-70018 (Centre of Excellence CELDIS).

\newpage
%\baselineskip=0.5\baselineskip
\listoffigures

\end{document}